\documentstyle[aps,pra,graphics,exscale,flafter,twocolumn]{revtex}
\newcommand{\beq}{\begin{equation} }
\newcommand{\eeq}{\end{equation} }
\begin{document}
\author{ Michael Mussinger, Aldo Delgado and Gernot Alber}
\title{Error Avoiding Quantum Codes and Dynamical Stabilization of
Grover's Algorithm}
\address{Abteilung f\"ur Quantenphysik,
Universit\"at Ulm, D-89069 Ulm, Germany }
\maketitle
\begin{abstract}
An error avoiding quantum code is presented which is capable
of stabilizing Grover's quantum search algorithm against a particular
class of coherent errors. This error avoiding code consists of states
only which are factorizable in the computational basis. Furthermore,
its redundancy is smaller than the one which is achievable with
a general error correcting quantum code saturating the quantum
Hamming bound. The fact that this code consists of factorizable states
only may offer advantages for the implementation of 
quantum gates in the error free subspace.
\end{abstract}
\narrowtext
\section{Introduction}
According to a suggestion of Feynman \cite{Feynman} quantum systems 
are not only of interest for their own sake but they
might also serve for practical purposes. Thus they may be used for 
simulating other quantum systems which are
less convenient to handle  or they may be used for
solving computational problems more efficiently than 
by any other classical means.
Two well known examples demonstrating this latter point 
are Shor's factorization algorithm\cite{Shor}
and Grover's search algorithm \cite{Grover}.

Quantum systems which are capable of performing quantum algorithms
are called quantum computers.
So far several physical systems have been considered as potential candidates
for quantum computers, such as trapped ions \cite{Cirac1},
nuclear spins of molecules \cite{NMR} or in the context of cavity quantum
electrodynamics atoms interacting with a single mode of the radiation field
\cite{Cirac2}.
To describe the operation
of a quantum computer theoretically it is advantageous to refrain from a
detailed physical description of the particular quantum system involved.
Thus, analogous to the spirit of computer science
it is more useful to
concentrate on those particular aspects 
which are essential for
the performance of quantum computation. On this abstract level
a generic quantum computer consists of $m$ distinguishable smaller quantum
systems which are frequently chosen as two-level 
systems with basis states
$|1\rangle$ and $|0\rangle$, for example.
The quantum information which can be stored in
one of these two level systems is called a qubit. 
Thus the state space of a generic quantum computer is spanned by
the so called computational basis which consists of the corresponding
$2^m$ product states $|b_0\rangle = |0 \ldots 00\rangle$, 
$|b_1\rangle = |0 \ldots 01\rangle$, ...
$|b_{2^m}\rangle = |1 \ldots 11\rangle$.

A typical 
quantum computation proceeds in several steps. 
Firstly, the quantum computer is prepared
in an initial state. Secondly, a
certain sequence of unitary transformations is performed which
are called quantum gates and which usually entangle the $m$ qubits.
Thirdly, the
final result is measured. Typically
the solution of a particular computational problem is
obtained with a certain probability only.
A general
quantum algorithm takes advantage of an essential feature
of quantum theory, namely the interference between probability amplitudes
and the fact that the dimensionality $D$ of the state space of $m$
distinguishable qubits increases exponentially with the number of qubits,
i.e. $D=2^m$.
One of the best known quantum algorithms are
the already mentioned Shor algorithm \cite{Shor}
and Grover's search algorithm \cite{Grover,Boyer}. 
In this latter algorithm a particular sequence of quantum
gates enables one to find a specific item out of
an unsorted database much
faster than with any other known classical mean.
This quantum algorithm was already realized
experimentally for a small number of qubits 
\cite{Chuang}. 

One of the main practical problems one has to overcome in the implementation
of quantum algorithms are non-ideal performances of the quantum gates
\cite{nonideal}
involved or random environmental 
influences which both tend to affect the relevant quantum coherence.
To protect quantum computation against such errors
two major strategies have been proposed recently, namely 
quantum error correction
\cite{Zurek96,Steane1,Steane} and error avoiding quantum codes 
\cite{Zanardi,Zanardi1}. Quantum error correction rests on the assumption that
nothing is known about the physical origin of the errors affecting
the quantum computation.
Its methods may be considered as
an extension of classical error correction techniques to the
quantum domain\cite{Steane1,Steane}.
The approach of the error avoiding quantum codes is different. They rest on the
assumption that the physical origin of the errors which affect the
quantum computation is known. The main idea of this latter approach
is to encode the logical information in one of those
subspaces of the relevant Hilbert
space which is not affected by the physical interactions responsible
for the occurance of errors \cite{Zanardi,Zanardi1}.
Both theoretical approaches to error correction rest on the concept of
redundancy which is also fundamental for classical
error correcting codes\cite{Welsh}. 
Provided the physical origin of the errors affecting a quantum computation
is known it is expected that error avoiding codes offer more
effective means for stabilizing quantum algorithms. 
This expectation is based on two facts. Firstly, there is no need for control
measurements which  are an essential ingredient for any error correcting code. 
Secondly, usually a smaller number
of {\em physical} qubits is needed for the representation
of a given number of {\em logical} qubits.

In the subsequent discussion it is demonstrated that this is indeed the
case. By considering Grover's quantum search algorithm it is shown that
non-ideal perturbations may be corrected dynamically in an efficient
way with the help of an appropriate error avoiding quantum code. As
a particular example,
we discuss coherent errors which may arise 
from systematic detunings of the physical qubits of the quantum computer
from the frequency of the light pulses which realize the required
quantum gates. 
It is shown that the corresponding error avoiding quantum code with
the lowest degree of redundancy is more efficient in encoding quantum
information than the corresponding optimal error correcting code
which saturates the quantum Hamming bound. The proposed error avoiding
quantum code consists of states only which are factorizable in
the computational basis. In this respect it differs significantly
from the recently proposed error avoiding code of Ref.\cite{Zanardi},
for example,
which also involves entangled states. Such factorizable codes may offer
practical advantages as far as the implementation of quantum gates
in error avoiding subspaces is concerned. 

The article is organized as follows: In Sec. II basic facts about
Grover's quantum search algorithm are summarized. It is demonstrated
that for large databases the dynamics of this quantum algorithm can
be described by a two-level Hamiltonian which implies Rabi oscillations
between the initial state and
the search state. In Sec. III general ideas underlying the construction
of error avoiding quantum codes are discussed. An efficient error avoiding
quantum code is presented which is capable of stabilizing Grover's
algorithm against a particular class of coherent errors. The redundancy
of this code is discussed and compared with the one resulting
from error correcting codes which saturate the quantum Hamming bound.
Numerical examples demonstrating the stabilizing capabilities
of this error avoiding quantum code
are presented in Sec. IV.
\begin{center}
\begin{figure}
\fbox
{
\begin{minipage}{7.5cm}
\vspace{0.2cm}
\begin{center}
\begin{minipage}{4.0cm}
\resizebox{3cm}{!}{\includegraphics{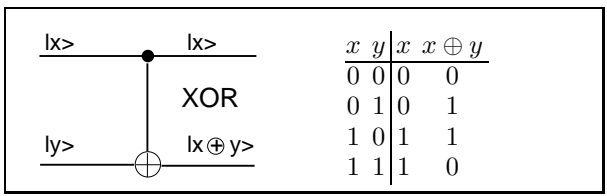}}
\end{minipage}
\begin{minipage}{3.0cm}
\begin{tabular}{cc|cc} 
$x$ & $y$ & $x$ & $x \oplus y$ \\ \hline
0 & 0 & 0 & 0 \\
0 & 1 & 0 & 1 \\
1 & 0 & 1 & 1 \\
1 & 1 & 1 & 0 
\end{tabular}
\end{minipage}
\end{center}
\end{minipage}
}
\vspace*{0.2cm}
\noindent
\caption{Quantum mechanical version of the classical-XOR gate as an
example for a quantum gate (CNOT gate):
The input state $|x,y\rangle$ is mapped onto the output state
$|x,x\oplus y\rangle$.\label{XOR}}
\end{figure}
\end{center}

\section{Grover's quantum search algorithm}
Consider an unsorted database with $N$ items
and a certain item $x_0$ you are searching for. 
As a particular
example you can imagine a
telephone directory with $N$ entries and a particular telephone number
$x_0$ you are looking for.
Furthermore, assume that you are given 
a black box, i.e. a so called oracle, which can decide whether an
item is $x_0$ or not.
Thus, in mathematical 
terms you are given a Boolean
function
\begin{equation}
f(x) = \delta_{x,x_0} = 
\left\{ \begin{array}{cc}
1 & x = x_0 \\
0 & x \neq x_0 
\end{array} \right.
\label{Boole}
\end{equation}
with $\delta_{a,b}$ denoting the Kronecker delta function. Usually the
elements $x$ of the 
database are assumed to be described by the $N$ integers between zero and
$N-1$.
Assuming that each application of the oracle requires one elementary
step 
a classical random search process  will require $N-1$ steps 
in the worst
case and one step 
in the best possible case. Thus, on the average a classical algorithm
will need $N/2$ steps to
find the searched item $x_0$.
It has been shown by
Grover \cite{Grover} that with the help of his
quantum search algorithm this task can be performed
in $O(\sqrt{N})$
steps with a probability arbitrarily close to unity.
The basic idea of this quantum algorithm
is to rotate the initial state 
of the quantum computation in the direction of the 
searched state $|x_0\rangle$ by a sequence of unitary
quantum versions of the oracle.
It
will become apparent from the subsequent discussion that
apart from Hadamard transformations the dynamics of this
rotation 
are analogous to a Rabi oscillation between the initially prepared state 
and the searched state $|x_0\rangle$.
It has been shown by Zalka \cite{Zalka} that Grover's quantum search
algorithm is optimal.

\subsection{Characteristic gate sequence of Grover's search algorithm} 
In Grover's quantum search algorithm 
every element of the database is represented by a state of the
computational basis of the quantum computer.
Thus a database which is represented by
$m$ qubits has $N=2^m$ distinguishable elements.
The state $|0..0110..0\rangle$ of the computational basis,
for example, corresponds to the element $0..0110..0$ of the
database in binary notation.
The quantum oracle ${\cal U}_f$ 
is determined completely by the Boolean function of Eq.(\ref{Boole})
and is represented by
a quantum gate, i.e. by the unitary and hermitian transformation
\begin{equation}
{\cal U}_f: |x,a\rangle \to |x, f(x) \oplus a\rangle.
\label{fx}
\end{equation}
Thereby $|x\rangle$ is an arbitrary element of the computational basis
and $|a\rangle$ is the state of an additional ancilla qubit which is 
discarded later.
The symbol $\oplus$ denotes
addition modulo 2.
This unitary form of the
oracle depends on the Boolean function $f(x)$.
As far as complexity estimates are concerned
it is
assumed that this unitary transformation requires
one elementary step.
This assumption is analogous to the complexity estimate of the
corresponding classical version of this search problem.

\newpage
\noindent
\begin{center}
\begin{figure}
\resizebox{6.5cm}{!}{\includegraphics{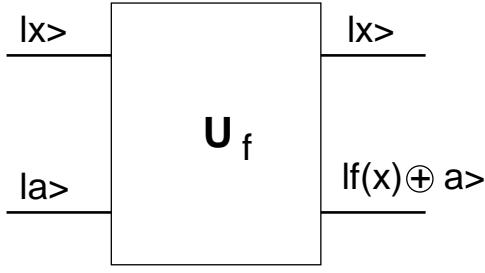}}
\vspace{0.2cm}
\noindent
\caption{Schematic representation of the quantum oracle ${\cal U}_f$:
For $f(x)\equiv x$ this quantum gate reduces to the CNOT gate of Fig.\ref{XOR};
for $|a\rangle \equiv |a_0\rangle = 1/\sqrt{2} \; (|0\rangle - |1\rangle )$
it results in the conditional phase inversion
$I_{x_0}$ of Eq.(\ref{formal}) needed in Grover's algorithm.
\label{Uf}}
\end{figure}
\end{center}

For the subsequent discussion it is important to note that
the elementary rotations in the direction of the searched quantum state 
$|x_0\rangle$ which are the key ingredient in Grover's algorithm
can be performed with the help of this unitary oracle.
Thus such a rotation can be performed without explicit knowledge of the
state $|x_0\rangle$. Its implicit knowledge
through the values of the Boolean function $f(x)$ is already sufficient. 
For large values of $N$ it turns out that the number of elementary
rotations needed to prepare state
$|x_0\rangle$ is $O(\sqrt{N})$. 
To implement such an elementary rotation from the initial state
$|s\rangle = |0...0\rangle$, for example, towards the final state
$|x_0\rangle$ two different types of quantum
gates are needed, namely 
{\em Hadamard} gates and {\em controlled phase inversions}.

A {\em Hadamard} gate is a unitary
one-qubit operation. 
It produces an equal weighted superposition of 
the two basis states according to the rule
\begin{eqnarray}
|0\rangle & \rightarrow & \frac{1}{\sqrt{2}} ( |0\rangle + |1 \rangle), \\
|1\rangle & \rightarrow & \frac{1}{\sqrt{2}} ( |0\rangle - |1 \rangle ) 
\end{eqnarray}
or in matrix notation
\[ H^{(2)} = \frac{1}{\sqrt{2}} \left( \begin{array}{cc} 
1 & 1 \\
1 & -1 
\end{array} \right).
\]
A $m$-qubit Hadamard gate $H^{(2^m)}$ is defined by the $m$-fold tensor product,
i.e. $H^{(2^m)}=H^{(2)}\otimes ...\otimes H^{(2)}$.
Thus, for two qubits, for example, $H^{(2^2)}$ is represented by the matrix
\begin{equation}
 H^{(2^2)} = \frac{1}{2} \left( \begin{array}{cccccccc} 
1 & 1 & 1 & 1 \\
1 & -1 & 1 & -1 \\
1 & 1 & -1 & -1 \\
1 & -1 & -1 & 1 
\end{array} \right).
\end{equation}
The Hadamard transformation is hermitian and unitary.
An arbitrary matrix element $H^{(2^m)}_{i,j}$
of a Hadamard transformation  may be written in the general form
\begin{equation}
 H^{(2^m)}_{i,j} = \frac{1}{\sqrt{2^m}} (-1)^{i \odot j}. 
\end{equation}
Thereby $i$ and $j$ denote binary numbers
and the multiplication $\odot$ is 
bitwise modulo 2, i.e. for $i=1$, $j=3$ and $m=2$, one obtains
$H^{(4)}_{1,3} = (1/2) 
(-1)^{( 01 \odot
11 )} =  (1/2) ( -1) ^ {( 0\cdot 1 + 1 \cdot 1 )} = - 1/2 $.
It has been shown by Grover \cite{Grover} that this Hadamard transformation
can be replaced by any other unitary one-qubit operation.

The remaining quantum gates needed for the implementation of the
necessary rotation are
{\em controlled phase inversions}  with respect to the initial
and searched states
$|s\rangle = |0...0\rangle$ and 
$|x_0\rangle$.
A controlled phase inversion with respect to a state $|x\rangle$
changes the phase of this particular state by an amount of $\pi$ and leaves all
other states unchanged.
Thus
the phase inversion $I_s$ with respect to the initial state $|s\rangle$
is defined by 
\begin{eqnarray}
I_s |s\rangle
&=& - |s\rangle,\nonumber\\
I_s |x\rangle &=& |x\rangle \hspace*{0.5cm}(x \neq s ).
\end{eqnarray}
For two qubits, for example, its matrix representation is given by
\begin{equation}
 I_s = \left( \begin{array}{cccc} 
- 1 & 0 & 0 & 0 \\
0 & 1 & 0 & 0  \\
0 & 0 & 1 & 0 \\
0 & 0 & 0 & 1 
\end{array} \right). 
\end{equation}
The controlled phase inversion $I_{x_0}$ with
respect to the searched state 
$|x_0\rangle$
is defined in an analogous way.
As state $|x_0\rangle$
is not known explicitly but only implicitly through the property
$f(x_0)=1$ this transformation has to be performed with
the help of the quantum oracle. 
This task can be achieved by preparing the ancilla
of the oracle of Eq.(\ref{fx})
in state
$ |a_0\rangle = 1/\sqrt{2} ( |0\rangle - |1\rangle )$. 
 As a consequence
one obtains the required properties for the phase inversion $I_{x_0}$, namely
\begin{eqnarray}
\lefteqn{|x,f(x) \oplus a_0 \rangle  \equiv
|x, 0 \oplus a_0 \rangle  \label{formal}} \\ \nonumber
&  & \hspace{3em} =  1/\sqrt{2} (|x,0\rangle - |x,1\rangle) =
|x,a_0\rangle  \;\; {\rm for }\;\; x \neq x_0 ,\\  \nonumber
\lefteqn{|x, f(x) \oplus a_0 \rangle \equiv
|x, 1 \oplus a_0 \rangle}  \\ \nonumber
& & \hspace{3em} = 1/\sqrt{2} (|x,1\rangle - |x,0\rangle) =
- |x,a_0\rangle \;\; {\rm for} \;\; x =x_0 .
\end{eqnarray}  
One should bear in mind that this controlled phase inversion can be
performed with the help of the quantum oracle of Eq.(\ref{fx})
only without explicit knowledge of state $|x_0\rangle$.

\begin{center}
\begin{figure}
\resizebox{4cm}{!}{\includegraphics{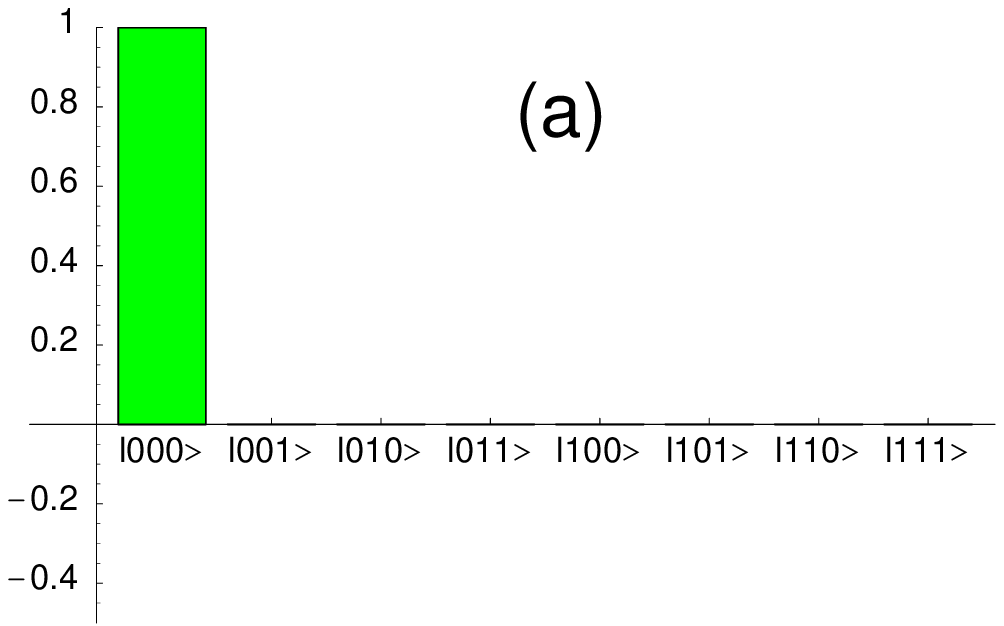}}
\resizebox{4cm}{!}{\includegraphics{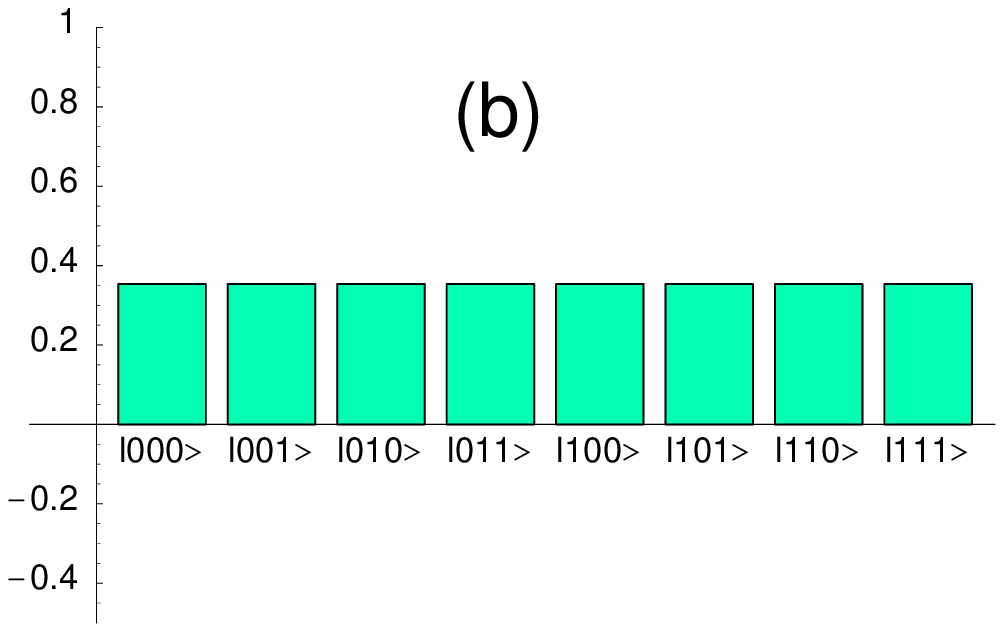}}

\noindent
\resizebox{4cm}{!}{\includegraphics{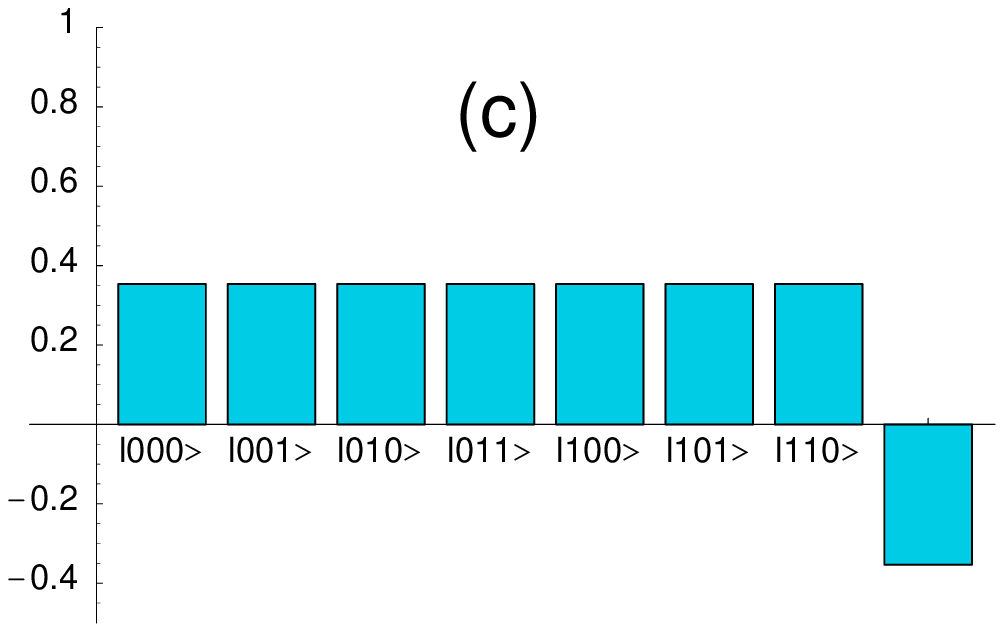}}
\resizebox{4cm}{!}{\includegraphics{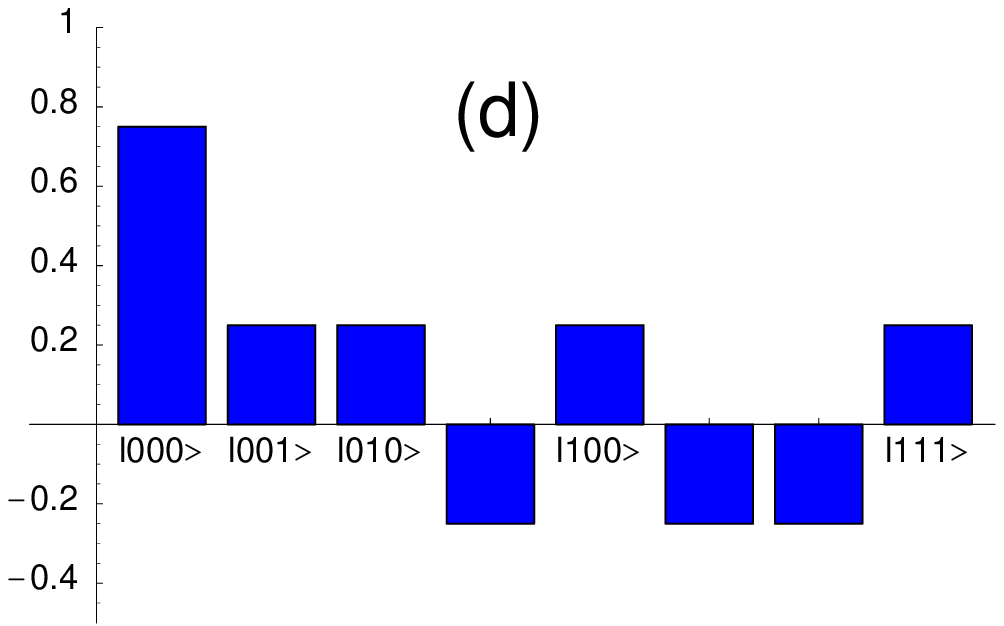}}

\noindent
\resizebox{4cm}{!}{\includegraphics{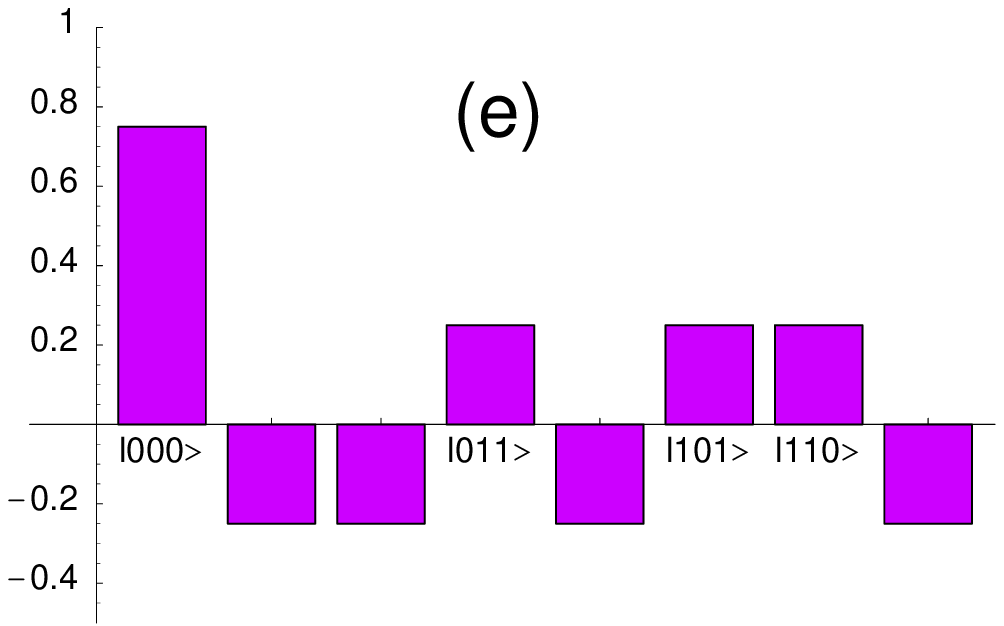}}
\resizebox{4cm}{!}{\includegraphics{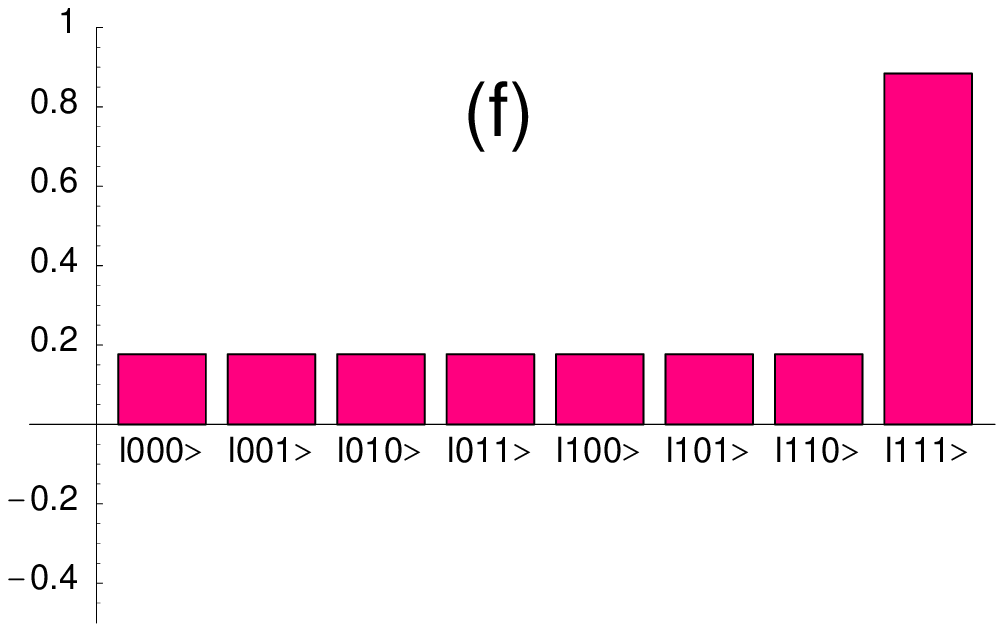}}
\vspace{0.2cm}
\noindent
\caption{Amplitude distributions 
resulting from the various quantum gates
involved in Grover's quantum search algorithm for the case
of three qubits: The quantum states which are prepared by these
gates are 
(a) $|s\rangle = |000\rangle$,
(b) $H^{(2^m)}|s\rangle$,
(c) $I_{x_0}H^{(2^m)}|s\rangle$,
(d) $H^{(2^m)}I_{x_0}H^{(2^m)}|s\rangle$,
(e) $-I_s H^{(2^m)}I_{x_0}H^{(2^m)}|s\rangle$,
(f) $-H^{(2^m)}I_s H^{(2^m)}I_{x_0}H^{(2^m)}|s\rangle$.
The searched state $|x_0\rangle$ entering
the Boolean function of Eq.(\ref{Boole}) is assumed to be state
$|111\rangle$. 
\label{grov_zus1}}
\end{figure}
\end{center}

Grover's algorithm starts 
by preparing all $m$ qubits of the quantum computer in state
$|s\rangle = |0...0\rangle$.
An elementary rotation in the direction of the searched state
$|x_0\rangle$ with the property $f(x_0)=1$ is achieved by the
gate sequence
\begin{equation}
Q = - I_{s} \cdot H^{(2^m)} \cdot I_{x_0} \cdot H^{(2^m)}.
\label{Q}
\end{equation}
In order to rotate the initial state $|s\rangle$ into state $|x_0\rangle$
one has to perform a sequence of $n$ such rotations and a final Hadamard
transformation at the end, i.e.
\begin{equation}
|f\rangle = H Q^n |s\rangle. 
\label{fop}
\end{equation}
The effect of the elementary rotation $Q$ is demonstrated in
Fig.\ref{grov_zus1}
for the case of three qubits, i.e. $m=3$.
The first Hadamard transformation $H^{(2^3)}$
prepares
an equally weighted state. The subsequent quantum gate
$I_{x_0}$ inverts the amplitude of the searched state
$|x_0\rangle = |111\rangle$.
Together with the subsequent Hadamard transformation and the phase
inversion $I_s$ this gate sequence $Q$ amplifies the probability amplitude
of the searched state $|111\rangle$. In this particular case
an additional Hadamard transformation finally prepares the quantum computer
in the searched state $|111\rangle$ with a probability of 0.88. 

In order to determine the dependence of the ideal number of repetitions
$n$ on the number of qubits $m$ it is convenient
to analyze the repeated application of the gate sequence $Q$
according to Eq.(\ref{fop})
in terms of the two states
$|s\rangle$ and $|v\rangle =H^{(2^m)}|x_0\rangle$ whose overlap is given by
$\epsilon = \langle s | v\rangle = \langle s | H^{(2^m)} |
x_0 \rangle = 2^{-m/2}$ for $m$ qubits.
It is straightforward to show that
the unitary gate sequence $Q$ preserves the
subspace spanned by these two states\cite{Grover}, i.e.
\begin{equation}
Q \left( \begin{array}{c}  |s\rangle \\
|v\rangle \end{array} \right) = 
\left(  \begin{array}{cc}
1 - 4 \epsilon^2 & 2 \epsilon \\
- 2 \epsilon & 1 \end{array} \right)  \left( \begin{array}{c}  |s\rangle \\
|v\rangle \end{array} \right). 
\end{equation}
Thus $Q$ acts like a rotation in the plane spanned by
states $|s\rangle$ and $|v\rangle$. The angle of rotation is
given by $\varphi = {\rm arcsin}(2\epsilon\sqrt{1-\epsilon^2})$.
\begin{center}
\begin{figure}
\resizebox{7.5cm}{!}{\includegraphics{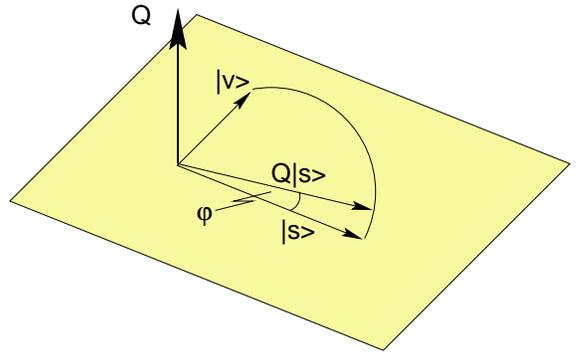}}
\vspace{0.2cm}
\noindent
\caption{Q is a rotation in the subspace spanned by states
$|s\rangle$ and $|v\rangle$.
\label{rotate}}
\end{figure}
\end{center}
After j iterations the amplitude of state $|v\rangle$ 
is given by
\cite{Boyer}
\begin{equation}
\sin \left[ (2j +1) \epsilon \right]. 
\end{equation}
Therefore, the optimal number $n$ of repetitions of the gate sequence $Q$
is approximately given by 
\begin{equation}
 n = \frac{\pi}{4 \;\; \arcsin \left( 2^{-m/2} \right) } - \frac{1}{2} \approx \frac{\pi}{4} 
\sqrt{2^m} \hspace*{0.2cm}(2^m\gg 1). 
\label{number}
\end{equation}

\subsection{Hamiltonian representation of Grover's algorithm}

If the database contains many elements, i.e. $N \equiv \epsilon^{-2}\gg 1$,
the repeated application
of the elementary rotation which is essential for Grover's search
algorithm can be described by a Hamiltonian quantum dynamics.
The elementary rotation $Q$ can be approximated by the relation
\begin{equation}
 Q = {\bf 1} -  \tau \; i/\hbar {\bf H}_G (\epsilon) + O(\epsilon^2) 
\end{equation}
which involves the Hamiltonian
\begin{equation}
{\bf H}_G =   2 i \epsilon \frac{\hbar}{\tau} \; \left(   
| v  \rangle  \langle s |  
- | s  \rangle  \langle v |
  \right). 
\label{hamiltonian}
\end{equation}
The elementary time $\tau$  might be interpreted as the
physical time  required for performing the elementary rotation $Q$.
The Hamiltonian of Eq.(\ref{hamiltonian}) describes the dynamics
of a quantum mechanical two level
system whose degenerate energy levels
$|s\rangle$ and $|v\rangle$ are coupled by a time-independent perturbation.
In lowest order of $\epsilon$ these degenerate energy levels are orthogonal.
The resulting oscillations between these coupled energy levels are
characterized by the Rabi frequency
$\Omega = 2 \langle  s | v \rangle / \tau$. 
Correspondingly,
the repeated application of the elementary rotation $Q$ can be determined
with the help of Trotter's product formula \cite{Trotter}, namely
\begin{eqnarray}
Q^n &=& (- I_s \cdot H^{(2^m)} \cdot  I_{x_0}  \cdot H^{(2^m)})^n
=\nonumber\\
&& \exp \left( - \frac{i}{\hbar}  
{\bf H}_G \; \cdot 
\tau n   \right) + O(\epsilon^2 n). 
\label{ntimes}
\end{eqnarray}
Thus, in the framework of this Hamiltonian description
applying the elementary rotation $Q$ $n$ times is equivalent to a
time evolution of the effective two-level quantum system over a time
interval of magnitude $n\tau$.
This Hamiltonian description demonstrates that the physics behind Grover's
quantum
search algorithm is the same as the physics governing the Rabi oscillations
between degenerate or resonantly coupled energy eigenstates.
As the errors entering Eq.(\ref{ntimes}) are of order $O(\epsilon^2 n)$
this Hamiltonian description is applicable only as long as
$\epsilon^2 n \equiv n/2^{m}\ll 1$. Thus for a given size of the
database it is valid only as long as
the number of iterations is sufficiently small, i.e. $n \ll 2^m$.
However, as Grover's search algorithm needs approximately
$(\pi\sqrt{2^{m}}/4)$ steps to find the searched item
the main condition which restricts the validity of 
this Hamiltonian description 
is a large size of the database, i.e. $\epsilon^2 \equiv 1/N \ll 1$.
\begin{center}
\begin{figure}
\resizebox{8.5cm}{!}{\includegraphics{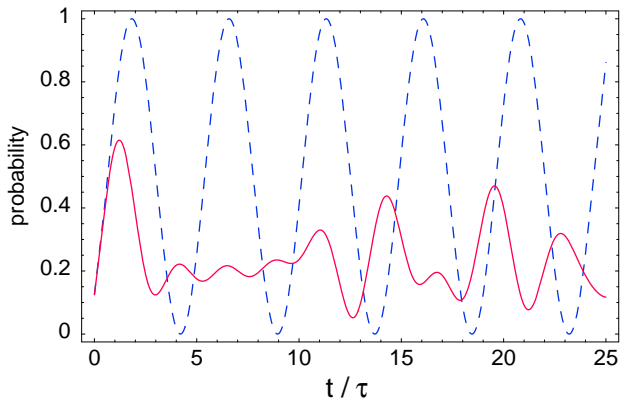}}
\vspace{0.2cm}
\noindent
\caption{The probability
of being in state $|x_0\rangle$  
after $n = t/\tau$
iterations of Grover's quantum search algorithm 
for three qubits:
the ideal dynamics according to the Hamiltonian time evolution
characterized by Eqs.(\ref{hamiltonian}) and (\ref{ntimes}) (dashed line);
the non-ideal case of coherent errors as characterized by Eqs.
(\ref{hamiltonian}), (\ref{ntimes}) and (\ref{pert}) (solid line)
with detunings
$\omega_1 = 0.5 \; \langle  s | v  \rangle / \tau$, 
$\omega_2 = 0.3  \; \langle  s | v  \rangle / \tau$, 
$\omega_3 = 0.2   \; \langle  s | v  \rangle / \tau$.
\label{detuning2}}
\end{figure}
\end{center}
\subsection{An example of coherent errors}
So far we have been concentrating on
the ideal dynamics of Grover's quantum search
algorithm. However, in practical applications it is very
difficult to realize
this search algorithm in an ideal way. Usually the ideal dynamics
are affected by numerous perturbations. 
Physically one may distinguish two different kinds of errors, namely
incoherent and coherent ones.
Typically
incoherent perturbations originate from a coupling of the physical
qubits of a quantum computer to an uncontrollable environment.
As a consequence the resulting errors are of a stochastic nature.
Coherent errors may arise from non-ideal 
quantum gates which lead to a unitary but non-ideal time evolution
of the quantum algorithm.
A simple example of this latter type of errors are systematic detunings
from resonance
of the light pulses with which the required quantum gates are realized
on the physical qubits.
In the Hamiltonian formulation of Grover's algorithm such
systematic detunings may be described by a perturbing Hamiltonian
of the form
\begin{equation}
{\bf H}_d = \sum_{i=1}^m  \hbar \omega_i \sigma_z^{(i)}.
\label{pert}
\end{equation}
In Eq.(\ref{pert}) it has been assumed that Grover's quantum algorithm
is realized by $m$ qubits and that the $i$-th qubit
is detuned with respect
to the ideal
transition frequency by an amount $\omega_i$. The 
Pauli spin-operator of the $i$-th qubit is denoted
$\sigma_z^{(i)}$.
In the presence of these systematic detunings
and for a large number of qubits the dynamics of Grover's algorithm
are described by the Hamiltonians of
Eqs.(\ref{hamiltonian}) and (\ref{pert}).

In order to obtain insight into the influence of this type of coherent
errors the performance of Grover's algorithm under
repeated applications of the elementary rotation $Q$ is depicted in Fig.
\ref{detuning2}. The dynamics of the ideal Grover algorithm are
depicted by the dashed line for the case of three qubits, i.e. $m = 3$.
The Rabi oscillations with frequency
$\Omega = 2\langle v|s\rangle /\tau$ are clearly visible.
The solid line shows the
probability of observing the quantum computer in state
$|x_0\rangle$ in a case in which all the qubits are detuned
from their ideal resonance frequency.
One notices the deviations from the ideal behaviour. Due to the
coherent nature of the errors the time evolution of the
non-ideal algorithm exhibits
revival phenomena \cite{Averbukh}.

\section{Error avoiding quantum codes}
In general there are two different strategies for correcting errors
in quantum information processing.
If nothing is known about the physical origin of the errors affecting
a qubit one can use {\em general quantum error correcting schemes}.
They may be viewed as
generalizations of classical error correction techniques to the
quantum domain \cite{Zurek96,Steane1,Steane}.
Typically they involve a suitably chosen quantum code and a sequence
of quantum measurements.
This code has to map all possible states
which may result from arbitrary environmental influences
onto orthogonal states.
According to basic postulates of quantum theory 
these orthogonal quantum states can be distinguished and based on the
result of a control measurement one may restore the 
original quantum state.
So far these general techniques have been applied mainly
to the stabilization of
static quantum memories \cite{Pellizzari}.
It is still an open question
whether these general methods
are 
also useful and efficient for the dynamical stabilization of quantum algorithms.

The second possible error correction strategy 
which seems to be well adopted also for stabilizing
quantum algorithms is based on {\em error avoiding quantum codes}
\cite{Zanardi,Zanardi1}. However, these latter
methods are applicable only, if the
physical origin of the errors is known.
The main idea is
to encode the quantum information in those subspaces of the Hilbert space
which are not affected by the errors.
This aim is achieved by restricting oneself to degenerate eigenspaces
of the relevant error operators. Thus, in the special case of
a single error operator, say ${\bf E}$, the basis states
$ \{ | \psi_i \rangle \}$ of such an error free subspace have to
fulfill the relation
\begin{equation}
{\bf E} | \psi_i \rangle = c \; | \psi_i \rangle.
\label{Eeigen}
\end{equation}
In the above mentioned example of
coherent errors which may affect Grover's algorithm this error operator
is given by the Hamiltonian of Eq.(\ref{pert}), i.e. ${\bf E} = {\bf H}_d$.
It is crucial for the success of an
error avoiding code that the eigenvalue $c$ of Eq.(\ref{Eeigen})
does not depend on the states belonging to the error free subspace.
This implies that all possible  elements of the error free subspace of the general
form $\sum_i \alpha_i |\psi_i\rangle$  are affected by the error operator in the
same way, i.e.
\beq {\bf E} (\sum_i \alpha_i  | \psi_i  \rangle) = c (\sum_i \alpha_i | 
\psi_i \rangle) .\eeq
It is apparent that
a non-trivial error avoiding code is possible only, if the eigenspace of 
the error operator ${\bf E}$ is
degenerate. 

\subsection{An error avoiding quantum code stabilizing coherent errors}
As an example for an error avoiding quantum code let us consider
the case of coherent errors which may affect Grover's quantum algorithm and
which can be characterized by the Hamiltonian ${\bf H}_d$ of Eq.(\ref{pert}).
In the simple case of equal detunings,
i.e. $\omega_1=...=\omega_m\equiv \omega$,
the error operator ${\bf E} $ reduces to the form
\begin{equation}
{\bf H}_e = \hbar \omega \sum_{i=1}^m  \sigma_z^{(i)}. 
\label{He}
\end{equation}
It is easy to find highly degenerate
error free subspaces of this error operator.
All states with a fixed number of ones and zeroes constitute a degenerate eigenspace
of ${\bf H}_e$.
For an even number of qubits it is  possible to find an error avoiding subspace
with eigenvalue $c=0$ so that 
\beq ({\bf H}_G + {\bf H}_e) \; 
| \psi \rangle = {\bf H} _G | \psi \rangle \eeq 
for all elements $|\psi\rangle$ of this subspace.
For this purpose one is looking for
quantum states with zero total spin.
For four qubits, for example, this subspace is defined by the basis vectors
$|0011 \rangle, |0101 \rangle, |0110
\rangle, |1001  \rangle, |1010 \rangle, |1100 \rangle$ 
and involves all states with the same number of zeros and ones.
Four of
these states may be used as a basis for the state space of two {\em logical} qubits.
For these eigenstates the error Hamiltonian ${\bf H}_e$ maps onto zero, e.g. 
\[ {\bf H}_e |0011\rangle = \hbar \omega \sum_{i=1}^{m=4} \sigma_z^{(i)}
|0011\rangle = \hbar \omega (1 + 1 - 1 - 1 ) | 0011\rangle  = 0. \] 
This particular error avoiding code works ideal for equal
detunings of all qubits from resonance.
It is formed by quantum states which factorize in the computational basis.
So it is expected that
in this error free subspace
the encoding 
of quantum information
and the implementation
of quantum gates is 
considerably easier than in cases in which the
error avoiding codes involve
entangled quantum states. 

\subsection{Implementation of quantum gates in an error free subspace}
To realize a quantum algorithm in an error free subspace one has to implement
the necessary quantum gates in such a way that they
do not mix the error free subspace with its orthogonal complement \cite{Bacon}.
Consider two logical qubits, for example,
which are encoded by four physical qubits. 
For this purpose one may choose the states 
$|0011 \rangle, |0101 \rangle, |0110
\rangle, |1001  \rangle$ which have been mentioned in the previous
subsection. 
This error avoiding code 
works ideal for stabilizing Grover's algorithm
with respect to the error operator ${\bf H}_e$ of Eq.(\ref{He})
provided it is possible to
realize the required unitary transformations, namely
Hadamard transformations and the controlled phase inversions.

Consider as an example a Hadamard transformation which 
acts in a two dimensional error avoiding subspace of this kind.
Thus it is assumed that the two basis states of this error avoiding code
are given by
$|01\rangle$  and $|10\rangle$ and that they involve two physical qubits.
Thus, we are looking for a transformation which performs the mappings
\begin{eqnarray}
|01\rangle &\to&
1/\sqrt{2} (|01\rangle + |10\rangle),\nonumber\\
|10\rangle &\to& 1/\sqrt{2}
(|01\rangle - |10\rangle)
\end{eqnarray}
and which does not mix the subspace spanned by $|01\rangle$ and
$|10\rangle$ with the orthogonal space spanned by the basis states
$|00\rangle$ and  $|11\rangle$. In matrix notation we are looking for a unitary
matrix of the form
\begin{equation}
 \left( \begin{array}{rrrr} 
\ast &  0 & 0 & \ast \\
0 & 1 & 1 & 0 \\
0 & 1 & - 1 & 0 \\
\ast & 0 & 0 & \ast 
\end{array} \right) 
\end{equation}
with $\ast$ denoting arbitrary entries which ensure unitarity.
Such a transformation can be achieved by the gate sequence
$CNOT_{21} ({\bf 1}\otimes \tilde{H}^{(2)})CNOT_{21}$ with $\tilde{H}^{(2)} = -
i \sigma_y H^{(2)}$. Thereby
$CNOT_{21}$ is a controlled-not operation with the first qubit as the
target and the second
qubit as the control qubit and $\sigma_y$ is the Pauli Matrix. 
Thus in matrix notation this relation yields
\begin{eqnarray}
&&\left( \begin{array}{cccc} 
1 &  0 & 0 & 0 \\
0 & 0 & 0 & 1 \\
0 & 0 &  1 & 0 \\
0 & 1 & 0 & 0  
\end{array} \right) 
\left( \begin{array}{rrrr} 
-1 &  1 & 0 & 0 \\
1 & 1 & 0 & 0 \\
0 & 0 &  -1 & 1 \\
0 & 0 & 1 & 1  
\end{array} \right)
\left( \begin{array}{cccc} 
1 &  0 & 0 & 0 \\
0 & 0 & 0 & 1 \\
0 & 0 &  1 & 0 \\
0 & 1 & 0 & 0  
\end{array} \right) =\nonumber\\
&&\left( \begin{array}{rrrr} 
-1 &  0 & 0 & 1 \\
0 & 1 & 1 & 0 \\
0 & 1 & - 1 & 0 \\
1 & 0 & 0 & 1 
\end{array} \right).
\end{eqnarray}
Obviously the final result does not mix the error avoiding subspace with its 
orthogonal complement.
But in the intermediate steps such a mixing takes place.
However, for practical purposes it is enough to ensure 
that the
time spent by the quantum computer in the orthogonal complement of the
error avoiding subspace is sufficiently small so that the resulting errors
can be neglected for all practical purposes.
Under these circumstances it is expected that the implementation of quantum
algorithms in error avoiding subspaces is a powerful tool for stabilizing quantum
codes.

\subsection{Code size of error avoiding quantum codes}
In order to estimate the redundancy which has to be introduced
for stabilizing a quantum algorithm
by an error avoiding quantum code let us consider the particular 
example of Sec. IIIA in more detail.
It has been argued that in the case of coherent errors which
can be characterized by the Hamiltonian of Eq.(\ref{He})
an error avoiding quantum code can
be constructed from basis states with equal numbers of ones and zeroes.
In order to minimize the redundancy it is desirable to maximize the
dimension of the resulting error avoiding subspace.
If one starts with $m$ physical qubits
the dimension $D(m,q)$ of the corresponding error avoiding
subspace with $q$ qubits in state $|1\rangle$ and $(m-q)$ qubits
in state $|0\rangle$, for example,
is given by 
\begin{equation}
D(m,q) = \left( \begin{array}{c} m \\ q \end{array} \right) \equiv
\frac{m!}{q!(m-q)!}.  
\end{equation}
From elementary properties of binomial coefficients it is clear that
$D(m,q)$ is maximum for $q=m/2$. Thus for an even number of qubits $m$
the largest possible dimension of the resulting error avoiding subspace
is given by
\begin{equation}
D(m,m/2) = \frac{m!}{[(m/2)!]^2} \to 2^m\sqrt{\frac{2}{m\pi}} \hspace*{0.2cm}(m\gg 1).
\end{equation}
Thus, in this case it is possible to encode 
\begin{eqnarray}
 l &=& {\rm log}_2 D(m,m/2) \to \nonumber\\
&&  m - \frac{{\rm log}_2 m}{2}
+{\rm log}_2\sqrt{2/\pi}
\hspace*{0.2cm}(m\gg 1) 
\label{maximum}
\end{eqnarray}
logical qubits with $m$ physical ones. 
It is instructive to compare the redundancy of this error 
avoiding code 
as described by Eq.(\ref{maximum})
with the ones resulting from general error correcting quantum codes
which saturate the quantum Hamming bound
\cite{Welsh,Gottesman}.
If one wants to correct arbitrary errors of maximum length $t$ 
with a general error correcting quantum code
the number of physical and logical qubits $m$ and $l$
have to fulfill
the so called quantum Hamming bound\cite{Zurek96,Welsh,Gottesman}, i.e.
\begin{equation}
2^l \sum_{r=0}^{t}3^r \left( \begin{array}{c}
m \\ r \end{array} \right) \leq 2^m.
\label{Hamming}
\end{equation}
This inequality reflects the fact that in a general error correcting
quantum code the action of different error operators onto
any of the logical qubits must lead to orthogonal quantum states.
The dimension of the resulting Hilbert space as described
by the left hand side of the inequality (\ref{Hamming})
has to be smaller than the
dimension of the Hilbert space of all physical qubits.
Thus the number of logical qubits obtainable by a general error
correcting code which is capable of correcting all possible errors 
of maximum length one, i.e. $t=1$,
cannot be larger than 
\begin{equation}
l_> = m - {\log}_2(3m+1). 
\label{Hammax}
\end{equation}
Comparing Eq.(\ref{maximum}) with Eq.(\ref{Hammax})
one realizes that the redundancy of this particular
error avoiding quantum code is smaller than the one
resulting from saturating the Hamming bound with a general error correcting code
capable of correcting errors of maximum length one, i.e. 
\begin{equation}
l - l_> \to \frac{1}{2}{\rm log}_2 m + {\rm log}_2\frac{3\sqrt{2}}{\sqrt{\pi}} > 0
\hspace*{0.2cm}(m\gg 1).
\end{equation}
However, this reduction of redundancy is based on the fact that
the error avoiding code obeying Eq.(\ref{maximum})
can stabilize errors only which are described by the Hamiltonian of
Eq.(\ref{He}).
Usually more general errors cannot be corrected with this code.
\begin{center}
\begin{figure}
\resizebox{8.5cm}{!}{\includegraphics{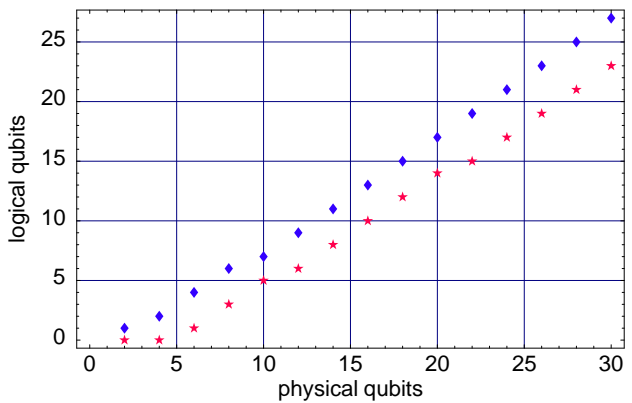}}
\vspace{0.2cm}
\noindent
\caption{Maximum number of logical qubits $l$ versus number of physical qubits $m$ for the 
error avoiding quantum codes which are capable of stabilizing the error operator of
Eq.(\ref{He}) (diamonds) (compare with Eq.(\ref{maximum})).
The corresponding relation $l_>(m)$ of Eq.(\ref{Hammax})
characterizing
the quantum Hamming bound is indicated by stars.
\label{code1_dim}}
\end{figure}
\end{center}

\section{Numerical examples}
In the previous section we have developed an error avoiding quantum code
which is capable of correcting coherent errors. These errors were
assumed to be caused by
systematic detunings of the physical qubits of the quantum computer from
the frequency of the laser pulses implementing the action of the quantum gates.
This error avoiding quantum code works perfect provided
all physical qubits are detuned from the frequency of these laser pulses
by the same amount. However, in realistic situations
this case is scarcely realized.
For the realistic assumption of unequal detunings in general the eigenstates of 
${\bf H}_d$ are non-degenerate so that it is not possible to construct
a perfect error avoiding quantum code.
Therefore the practical question arises whether the presented
error avoiding quantum code
of Sec. III is still useful for stabilizing quantum
algorithms against arbitrary systematic detunings.
\begin{center}
\begin{figure}
\resizebox{8.5cm}{!}{\includegraphics{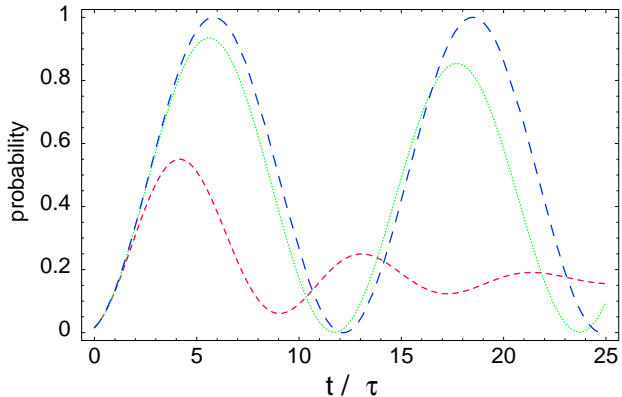}}
\vspace{0.2cm}
\noindent
\caption{Probability of finding the quantum computer in the searched state
$|x_0\rangle$ after $n=t/\tau$ iterations:
ideal dynamics without detunings for 6 qubits(dashed line), 
with detunings and without error avoiding encoding for 8 qubits(dotted line),
with detunings and with error avoiding encoding using 8 physical
qubits which can encode the quantum information of 6 logical qubits
(solid line). 
For the latter two cases the magnitude of the detunings $\omega_i$
of the 8 qubits which determine the error operator of Eq.(\ref{pert})
are given by
$\boldmath\omega_i \tau/\langle v|s\rangle = 0.92065$, $1.1436$, 
$0.71449$, $1.39566$, $1.29707$, $0.70149$, $1.19195$, $1.00343$. 
\label{detuning}}
\end{figure}
\end{center}

The dynamics of Grover's algorithm in the presence of 
arbitrary detunings are depicted in Fig. \ref{detuning}.
The dashed line represents the ideal dynamics in the absence
of detunings for the case of 6 qubits as evaluated from the
Hamiltonian of Eq.(\ref{hamiltonian}). The characteristic
Rabi oscillations are clearly apparent. The corresponding
dynamics for 8 qubits in the
presence of arbitrarily chosen detunings are depicted by the
dotted line in Fig.\ref{detuning}. It is apparent that 
in this case a quantum search for state $|x_0\rangle$
is not successful at all.
However, as apparent from the solid line of Fig. \ref{detuning}
encoding the quantum information by the error avoiding
code of Sec.III improves the performance considerably.
Despite the fact that this error avoiding code has not been
designed for these detunings it almost succeeds
in finding the searched quantum state $|x_0\rangle$ after a number
of iterations which is close to the ideal case 
(compare with Eq.(\ref{number})).

\begin{center}
\begin{figure}
\resizebox{8.5cm}{!}{\includegraphics{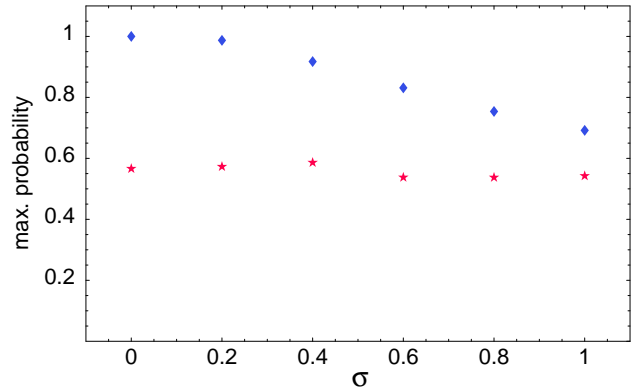}} 
\vspace{0.2cm}
\noindent
\caption{Average maximum probability of success for Grover's algorithm
with 8 qubits in the presence of randomly chosen detunings:
with error avoiding encoding according to Sec. III (diamonds);
without error avoiding encoding (stars).
The detunings $\omega_i$ of the 8 physical
qubits were chosen randomly according to a
normal distribution with mean value
$\overline{\omega}=0.5\langle v|s\rangle/\tau$. The corresponding
variance $\sigma$ of these
detunings is plotted on the x-axis in units of the mean value
$\overline{\omega}$. 
\label{det_stat} }
\end{figure}
\end{center}

In order to obtain more insight into the stabilizing properties of this
error avoiding code let us investigate the probability of
success in the presence of arbitrary detunings in more detail.
For this purpose we consider 8 physical qubits whose detunings $\omega_i$ are
distributed randomly according to a normal distribution.
According to Fig.\ref{code1_dim} these 8 physical qubits
are capable of encoding 6 logical qubits.
In Fig. \ref{det_stat} the average value of the
maximum probability of finding the quantum computer in the searched
state $|x_0\rangle$ is depicted for various values of the variance
of the randomly chosen detunings. 
The lower sequence of dots (stars) refers to Grover's algorithms without
error avoiding encoding and the upper sequence of points (diamonds) refers
to error avoiding encoding according to Sec. III. It is apparent that
error avoiding encoding is very successful as long as the differences
between the detunings of the qubits is sufficiently small.
Only in extreme cases in which these differences become comparable
to the typical magnitudes of the detunings this type this error
avoiding code is no longer capable of stabilizing Grover's algorithm
in a satisfactory way.

\section{Summary and Conclusions}
It has been demonstrated
that error avoiding quantum codes may offer efficient methods
for stabilizing quantum codes dynamically against those types of
errors whose physical origin is known. As a particular example
we discussed the stabilization of Grover's quantum search algorithm
against coherent errors which may arise from systematic detunings
of the physical qubits from the frequency of the light pulses
implementing the quantum gates. Though originally the error
avoiding quantum code has been
constructed for the special case of equal detunings of all the qubits
it has been shown that it is also capable
of stabilizing this quantum algorithm in other cases to a satisfactory
degree. The proposed error avoiding quantum code consists of quantum
states only which are factorizable in the computational basis.
This may offer advantages as far as the implementation of the necessary
quantum gates in this error free subspace is concerned. Furthermore,
this quantum code has also other noteworthy properties, such as a
redundancy which is lower than the one of an optimal error correcting
quantum code saturating the quantum Hamming bound.
Though 
the stabilizing ability of error avoiding quantum
codes has been demonstrated for one particular quantum code and
one particular class of coherent errors only it is expected that
similar capabilities are also found in
more general cases which may also involve incoherent
errors.
\\
\\
This work is supported by the DFG within the SPP
`Quanteninformationsverarbeitung'. Stimulating discussions with Thomas
Beth, Markus Grassl and Dominik Janzing
are acknowledged.

\end{document}